# Integrating Information Technology in Healthcare: Recent Developments, Challenges, and Future-Prospects for Urban and Regional Health

**Shipu Debnath[1*]**

[1]Department of Urban And Regional Planning, Jahangirnagar University, Dhaka, Bangladesh

[*]Corresponding Author E-mail: Debnathshipu@gmail.com

**Abstract**

The use of technology in healthcare has become increasingly popular in recent years, with the potential to improve how healthcare is delivered, patient outcomes, and cost-effectiveness. This review paper provides an overview of how technology has been used in healthcare, particularly in cities and for personalized medicine.

The paper discusses different ways technology is being used in healthcare, such as electronic health records, telemedicine, remote monitoring, medical imaging, wearable devices, and artificial intelligence. It also looks at the challenges and problems that come with using technology in healthcare, such as keeping patient data private and secure, making sure different technology systems can work together, and ensuring patients are comfortable using technology.

In addition, the paper explores the potential of technology in healthcare, including improving how easily patients can get care, the quality of care they receive, and the cost of care. It also talks about how technology can help personalize care to individual patients. Finally, the paper summarizes the main points, makes recommendations for healthcare providers and policymakers, and suggests directions for future research. Overall, this review shows how technology can be used to improve healthcare, while also acknowledging the challenges that come with using technology in this way.

# 1. Introduction

Healthcare is a vital aspect of human society, and its improvement has always been a major priority for policymakers, researchers, and practitioners. In recent years, there has been a significant increase in the use of information technology (IT) in healthcare to improve patient outcomes and quality of care [1]. Information technology encompasses a wide range of tools and systems, including electronic health records (EHRs), telemedicine, medical imaging, mobile health, and artificial intelligence (AI), among others.

The integration of IT in healthcare has shown promise in facilitating more efficient and effective healthcare delivery. For instance, EHRs and clinical information systems have been shown to improve clinical decision-making, reduce medical errors, and enhance patient safety. Telemedicine and remote monitoring have increased access to healthcare services, particularly for underserved populations and those in remote areas. Medical imaging and diagnostic tools have improved accuracy and speed of diagnosis, leading to better treatment outcomes [2]. Mobile health and wearable devices have facilitated patient self-management and monitoring, enabling better disease management [3]. AI and machine learning have shown potential in predicting and diagnosing diseases, identifying patterns, and optimizing treatment plans [4].

Despite these advancements, the integration of IT in healthcare also faces significant challenges and limitations. Privacy and security concerns have been raised about the use of EHRs and other healthcare IT systems. Interoperability and standardization of healthcare IT systems are also major challenges, as they impact the exchange of information across different healthcare providers and systems [5]. Additionally, human factors such as user acceptance and training have been identified as important factors that affect the successful implementation and adoption of healthcare IT. Legal and regulatory issues related to data sharing, liability, and reimbursement also pose challenges. Moreover, ethical and social implications of using AI and machine learning in healthcare, such as bias and discrimination, require careful consideration [6].

The potential of IT in healthcare is significant, and it has the potential to revolutionize healthcare delivery in terms of access, quality, and cost-effectiveness. Improved access and quality of care, enhanced patient engagement and self-management, increased efficiency and cost-effectiveness, and advanced analytics and personalized medicine are some of the potential benefits of IT in healthcare [7].

The objective of this review is to explore the recent developments, challenges, and future prospects of integrating IT in healthcare. The review will identify and summarize recent applications of IT in healthcare, explore the challenges and limitations of integrating IT in healthcare, discuss the potential of IT in healthcare, and outline future prospects and opportunities for research and practice.

## 2. Recent Developments and Applications of Information Technology in Healthcare

### A. Electronic Health Records and Clinical Information Systems

Electronic health records (EHRs) and clinical information systems (CIS) are essential tools that have transformed healthcare delivery by digitizing and streamlining patient records and clinical workflows [8]. EHRs are digital records of patients' health information, including medical history, diagnoses, medications, and test results, among other information. CIS are software systems that support clinical workflows, including order entry, medication administration, and results management, among others [9].

The use of EHRs and CIS has been shown to improve clinical decision-making, enhance patient safety, and reduce medical errors. They also facilitate communication and coordination among healthcare providers and support continuity of care. EHRs and CIS can also generate alerts and reminders for healthcare providers, enabling them to make more informed clinical decisions [10].

Moreover, the use of EHRs and CIS has shown potential in improving healthcare quality and patient outcomes. For example, a study found that hospitals that implemented EHRs had lower rates of adverse events, shorter lengths of stay, and lower mortality rates compared to those that did not [11]. Another study showed that the use of a CIS was associated with improved adherence to clinical guidelines and better outcomes for patients with acute myocardial infarction [12].

However, the integration of EHRs and CIS in healthcare also poses challenges. One of the main challenges is the issue of interoperability and data sharing among different healthcare providers and systems. Another challenge is the cost and time required to implement and maintain EHRs and CIS. Moreover, the use of EHRs and CIS raises concerns about data privacy and security, as patient data is stored electronically and is vulnerable to cyber threats.

In conclusion, the use of EHRs and CIS has transformed healthcare delivery by digitizing and streamlining patient records and clinical workflows. The benefits of using EHRs and CIS include improved clinical decision-making, enhanced patient safety, and improved healthcare quality and patient outcomes. However, the integration of EHRs and CIS also poses challenges, including interoperability, cost, and data privacy and security.

### B. Artificial Intelligence and Machine Learning

Artificial intelligence (AI) and machine learning have the potential to transform healthcare by improving community health, remote care service, early detection, healthcare resource utilization [13]. AI algorithms can be trained on large datasets of patient information to identify patterns and make predictions about patient outcomes, which can help healthcare providers make more accurate diagnoses and personalized treatment plans [14].

One example of the use of AI in healthcare is the development of computer-aided diagnosis (CAD) systems, which use machine learning algorithms to analyze medical images and assist radiologists in identifying abnormalities [15]. Another example is the use of AI to analyze electronic health records (EHRs) to predict patient outcomes and identify patients at risk of developing certain conditions.

However, the use of AI in healthcare also raises ethical and privacy concerns. For example, there are concerns about the potential for AI algorithms to reinforce biases and discriminate against certain patient groups. Additionally, there are concerns about the security of patient data and the potential for unauthorized access or misuse of this information.

To address these concerns, it is important to develop guidelines and regulations for the ethical use of AI in healthcare, as well as to ensure that patients are informed about how their data will be used and protected [16]. With careful consideration and implementation, AI and machine learning have the potential to greatly improve healthcare outcomes and patient care.

### C. Telemedicine and Remote Monitoring

Telemedicine and telehealth have become increasingly popular in recent years, offering a range of benefits to both patients and healthcare providers. Telemedicine refers to the use of technology to provide clinical healthcare services from a distance, while telehealth is a broader term that includes the use of technology to support healthcare services and information [17].

Telemedicine and telehealth can improve access to healthcare services, particularly for patients in rural or remote areas. They can also reduce healthcare costs and improve patient outcomes. For example, a study found that telemedicine consultations for patients with diabetes resulted in significant improvements in HbA1c levels, blood pressure, and lipid levels [18].

Moreover, telemedicine and telehealth can enhance communication and collaboration among healthcare providers, allowing for more coordinated and efficient care. They can also improve patient engagement and self-management, as patients can use technology to monitor their health status and receive support from healthcare providers.

However, the use of telemedicine and telehealth also poses challenges. One of the main challenges is the issue of reimbursement, as not all telemedicine services are covered by insurance. Another challenge is the need for appropriate technology infrastructure and support to ensure the reliability and security of telemedicine services. Moreover, the use of telemedicine and telehealth raises concerns about the potential loss of human connection and empathy in healthcare delivery [19].

In conclusion, telemedicine and telehealth offer a range of benefits to both patients and healthcare providers, including improved access to healthcare services, reduced costs, and enhanced communication and collaboration. However, the use of telemedicine and telehealth also poses challenges, including reimbursement, technology infrastructure, and concerns about the loss of human connection in healthcare delivery.

### D. Medical Imaging and Diagnostic Tools

Medical imaging and diagnostic tools have revolutionized the diagnosis and treatment of various medical conditions. These tools include X-rays, computed tomography (CT) scans, magnetic resonance imaging (MRI), ultrasound, and positron emission tomography (PET) scans [20]. These imaging technologies use different types of energy to create images of the inside of the body, allowing healthcare providers to identify and diagnose a wide range of conditions.

For example, CT scans use X-rays to create detailed images of the body and are commonly used to diagnose conditions such as bone fractures, tumors, and blood clots [21]. MRIs use strong magnetic fields and radio waves to produce images of the body and are particularly useful in diagnosing conditions such as brain and spinal cord injuries, tumors, and multiple sclerosis [22].

Ultrasound imaging uses high-frequency sound waves to create images of internal organs and structures and is commonly used to monitor the health of fetuses during pregnancy and to diagnose conditions such as gallstones and thyroid nodules. PET scans use a radioactive tracer to create images of the body's metabolic activity, and are commonly used to diagnose cancer, heart disease, and neurological disorders.

Despite the many benefits of medical imaging and diagnostic tools, there are also concerns about their potential risks and limitations. For example, some imaging technologies expose patients to ionizing radiation, which can increase the risk of cancer [23]. Additionally, the high cost of some imaging technologies can be a barrier to access, particularly in low-income countries [24].

In conclusion, medical imaging and diagnostic tools have transformed the practice of medicine, allowing healthcare providers to diagnose and treat a wide range of conditions. However, these tools also pose risks and limitations that need to be carefully considered to ensure the safe and effective use of these technologies.

### E.  Mobile Health and Wearable Devices

Mobile health (mHealth) and wearable devices have transformed healthcare by providing patients with access to real-time health data and empowering them to manage their own health. These technologies include mobile apps, wearable sensors, and other wireless devices that can be used to monitor various health metrics such as heart rate, blood pressure, and sleep patterns. Mobile applications are using machine learning and deep learning algorithms in diseases detection by just clicking images. Early detection of skin diseases like melanoma, psoriasis, folliculitis helps their users to take immediate actions and preventing severe scenarios [25].

Moreover, mHealth technology is a mobile app that helps patients manage chronic conditions such as diabetes by allowing them to track their blood glucose levels and receive personalized recommendations for managing their condition [26]. Wearable sensors, such as fitness trackers and smartwatches, can monitor physical activity levels and provide feedback on exercise and movement habits, encouraging users to lead more active lifestyles [27].

Another potential benefit of mHealth and wearable devices is their ability to improve patient outcomes by providing more personalized and timely care. For example, wearable devices can alert healthcare providers to changes in a patient's vital signs or activity levels, allowing for early intervention and potentially preventing hospitalizations. Additionally, mHealth technologies can be used to facilitate remote consultations and telemedicine, enabling patients to access medical care from their homes or other remote locations.

Despite the potential benefits of mHealth and wearable devices, there are also concerns about the privacy and security of patient data. Patients must be informed about the potential risks and limitations of using these technologies, and healthcare providers must take steps to ensure that patient data is kept secure and confidential [28].

In conclusion, mobile health and wearable devices have revolutionized the way patients manage their health and interact with healthcare providers. These technologies have the potential to improve patient outcomes, promote healthy lifestyles, and increase access to medical care. However, it is important to carefully consider the risks and limitations of these technologies to ensure that patient data is kept safe and confidential.

**3. Challenges and Limitations of Integrating Information Technology in Healthcare:**

Despite the potential benefits of integrating information technology (IT) in healthcare, there are several challenges and limitations that need to be addressed. Some of the most significant challenges include the following:

A. **Data Privacy and Security**: One of the biggest concerns associated with the integration of IT in healthcare is the protection of patient data privacy and security [29]. As patient data is increasingly being digitized and shared across different systems and devices, it becomes more vulnerable to data breaches and cyberattacks.

B. **Interoperability and Standardization:** Another major challenge is the lack of interoperability and standardization across different healthcare IT systems and devices. The lack of standardization makes it difficult to share data and communicate effectively across different systems, which can lead to medical errors and poor patient outcomes.

C. **Human Factors and User Acceptance:** The successful implementation of IT in healthcare also depends on human factors, such as user acceptance and adoption [29]. Patients and healthcare providers may have varying levels of comfort and familiarity with IT systems, which can impact their willingness to use and adopt new technologies.

D. **Legal and Regulatory Issues:** The integration of IT in healthcare is also subject to various legal and regulatory requirements, such as data privacy laws and regulations [29]. Failure to comply with these requirements can result in legal and financial consequences for healthcare organizations.

E. **Ethical and Social Implications:** Lastly, the integration of IT in healthcare raises ethical and social implications that need to be carefully considered [30]. For instance, the use of AI and machine learning in healthcare decision-making may raise concerns about bias and discrimination.

Despite these challenges, there are promising developments in the field of healthcare IT, such as the use of mobile devices for skin disease detection and diabetes management, and the development of smartwatch apps for remote elderly monitoring and fitness tracking. Prospects for healthcare IT include the use of telemedicine and virtual reality technologies to improve access to healthcare services and enhance patient outcomes.

**4. Potential of Information Technology in Healthcare**

The integration of information technology (IT) in healthcare has the potential to transform the way healthcare is delivered and improve patient outcomes. Some of the potential benefits of IT in healthcare include the following:

A. **Improved Access and Quality of Care:** IT can help improve access to healthcare services, especially for underserved populations, by enabling remote consultations and telemedicine services. IT can also help improve the quality of care by providing real-time access to patient information and facilitating communication and collaboration among healthcare providers.

B. **Enhanced Patient Engagement and Self-Management:** IT can also enhance patient engagement and self-management by providing patients with access to their health information and enabling them to monitor their health status [31]. This can help patients take an active role in their healthcare and improve their overall health outcomes.

C. **Increased Efficiency and Cost-Effectiveness:** The use of IT in healthcare can also increase efficiency and cost-effectiveness by reducing administrative burdens and streamlining healthcare processes [32]. This can lead to significant cost savings for healthcare organizations and improve the overall quality of care.

D. **Advanced Analytics and Personalized Medicine:** IT can also enable advanced analytics and personalized medicine by leveraging big data and machine learning algorithms to analyze patient data and develop personalized treatment plans [33]. This can lead to more precise diagnoses and treatments and improve patient outcomes.

Despite these potential benefits, there are still several challenges and limitations that need to be addressed to fully realize the potential of IT in healthcare, as discussed in the previous section.

# 5. Future Prospects of Information Technology in Healthcare

### A. Emerging Technologies and Trends

In recent years, mobile technology has been harnessed to develop new healthcare technologies, including apps and devices that can monitor and track a wide range of health concerns. For example, mobile apps have been developed for skin disease detection, food dieting and diabetes management, fall detection in elderly people [34], and fitness tracking [35].

Smartwatches are also emerging as a promising healthcare technology, with capabilities including COVID-19 alert by detecting hand transition [36], water consumption detection, remote elderly monitoring [37], fitness tracking, and harmful activity monitoring in mental health patients [38].

### B. Policy and Regulatory Developments

As these technologies continue to evolve, there will be a need for policies and regulations to guide their use. This includes issues related to data privacy and security, interoperability and standardization, and ensuring that new technologies are developed and deployed in an equitable and socially responsible manner [39].

### C. Innovations in Business Models and Service Delivery

The integration of healthcare technology will also require innovations in business models and service delivery. This includes the development of new payment models that incentivize the use of technology to improve health outcomes, as well as new service delivery models that incorporate technology to provide more efficient and effective care [40].

### D. Opportunities and Challenges for Research and Practice

As technology continues to advance, there will be opportunities for research and practice to explore new ways of using technology to improve health outcomes. This will require collaboration between healthcare providers, technology developers, and researchers to develop and test new interventions. However, there are also challenges and limitations that need to be addressed, such as ensuring that new technologies are accessible and affordable to all, and addressing concerns related to user acceptance and ethical implications [41].

On the other hand, mobile phones and smartwatches have become essential tools for individuals to manage their health and wellness. With the proliferation of mobile health apps and wearable devices, people can now easily monitor and track various aspects of their health, from physical activity and diet to chronic conditions and mental health [42].

**Mobile:**

- Skin disease detection: Mobile apps that use artificial intelligence (AI) to analyze photos of skin lesions could improve early detection of skin cancer and other dermatological conditions [43].

- Food dieting/diabetes management: Mobile apps that help users track their food intake and blood sugar levels could improve diabetes management and prevent complications [44].

- Fall detection in elderly people: Mobile apps that use sensors to detect falls and alert caregivers or emergency services could help prevent injuries and save lives [45].

- Fitness applications: Mobile apps that provide exercise programs, track physical activity, and monitor vital signs could help users achieve their fitness goals and prevent chronic diseases [46].

**Smartwatch:**

- COVID alert by detecting hand transition: Smartwatches equipped with sensors that can detect hand gestures could help identify early symptoms of COVID-19, such as coughing and shortness of breath [47].

- Water consumption detection: Smartwatches that monitor water intake and remind users to drink more water could help prevent dehydration and promote healthy habits (87).

- Remote elderly monitoring: Smartwatches that track vital signs, medication adherence, and activity levels could help caregivers monitor elderly patients from a distance and detect health problems early.

- Fitness tracker: Smartwatches that track physical activity, sleep, and heart rate could help users stay motivated to exercise and maintain a healthy lifestyle.

- Harmful activity monitoring in mental patients: Smartwatches that use AI to detect signs of agitation or distress in mental health patients could help prevent self-harm or violent behavior [48].

# 6. Conclusion

## A. Summary of the main points

Mobile phones and smartwatches have become essential tools for individuals to manage their health and wellness. They provide convenient access to health information and can assist in the detection and management of various health conditions, such as skin diseases, diabetes, falls in elderly people, and mental health issues. Additionally, mobile health apps and wearable devices can promote physical activity and encourage healthy behaviors, such as adequate water consumption.

## B. Implications and recommendations for practice and policy

The widespread adoption of mobile health technologies has important implications for healthcare practice and policy [49]. Healthcare providers should be prepared to incorporate mobile health data into clinical decision-making and provide guidance to patients on the use of mobile health apps and wearable devices. Additionally, policymakers must consider regulatory and privacy issues related to the collection and use of health data by mobile devices.

## C. Limitations and future research directions

Despite the potential benefits of mobile health technologies, there are also limitations and challenges that must be addressed [49]. For example, there is a need for further research to assess the accuracy and reliability of mobile health apps and devices. Additionally, there are concerns regarding the potential for mobile health technologies to exacerbate health disparities [50]. Future research should also explore the use of mobile health technologies in underserved populations and in the context of global health.

## D. Final thoughts and reflections

Overall, mobile health technologies have the potential to transform healthcare delivery and improve health outcomes [49]. As technology continues to evolve and becomes more integrated into healthcare, it will be important for healthcare providers and policymakers to stay informed and adapt to these changes. With careful consideration of the benefits and limitations of mobile health technologies, we can work towards a future where healthcare is more accessible, efficient, and effective for all.

7. References:

1. Institute of Medicine (US) Committee on Quality of Health Care in America. (2001). Crossing the quality chasm: a new health system for the 21st century. National Academies Press (US).

2. Levin, D. C., & Rao, V. M. (2008). Turf wars in radiology: a potential threat to access, quality, and cost of imaging. Radiology, 247(3), 610-614.

3. Free, C., Phillips, G., Watson, L., Galli, L., Felix, L., Edwards, P., ... & Haines, A. (2013). The effectiveness of mobile-health technology-based health behaviour change or disease management interventions for health care consumers: a systematic review. PloS One, 8(1), e55589.

4. Obermeyer, Z., & Emanuel, E. J. (2016). Predicting the future—big data, machine learning, and clinical medicine. New England Journal of Medicine, 375(13), 1216-1219.

5. Office of the National Coordinator for Health Information Technology. (2015). Connecting health and care for the nation: a shared nationwide interoperability roadmap. US Department of Health and Human Services.

6. Price, W. N., & Cohen, I. G. (2019). Privacy in the age of medical big data. Nature Medicine, 25(1), 37-43.

7. World Health Organization. (2019). Digital health. Retrieved from https://www.who.int/health-topics/digital-health#tab=tab_1

8. Adler-Milstein, J., & Jha, A. K. (2014). HITECH Act drove large gains in hospital electronic health record adoption. Health Affairs, 33(7), 1416-1422.

9. Bates, D. W., & Gawande, A. A. (2003). Improving safety with information technology. New England Journal of Medicine, 348(25), 2526-2534.

10. Jones, S. S., Adams, J. L., Schneider, E. C., Ringel, J. S., McGlynn, E. A., & Electronic Health Record Adoption and Quality of Care in the Veterans Health Administration. Health Services Research, 49(1pt2), 436-452.

11. Amarasingham, R., Plantinga, L., Diener-West, M., Gaskin, D. J., & Powe, N. R. (2009). Clinical information technologies and inpatient outcomes: a multiple hospital study. Archives of Internal Medicine, 169(2), 108-114.

12. Poon, E. G., Wright, A., Simon, S. R., Jenter, C. A., Kaushal, R., Volk, L. A., & Bates, D. W. (2004). Relationship between use of electronic health record features and health care quality: results of a statewide survey. Medical Care, 42(9), 810-819.

13. Roy, Mrinmoy, et al. "Machine Learning Applications In Healthcare: The State Of Knowledge and Future Directions." arXiv preprint arXiv:2307.14067 (2023).

14. Obermeyer Z, Emanuel EJ. Predicting the Future - Big Data, Machine Learning, and Clinical Medicine. N Engl J Med. 2016 Sep 29;375(13):1216-9.

15. Park SH, Han K. Methodologic Guide for Evaluating Clinical Performance and Effect of Artificial Intelligence Technology for Medical Diagnosis and Prediction. Radiology. 2018 Oct;289(1):17-25.

16. European Commission. Artificial Intelligence: Ethics Guidelines for Trustworthy AI. 2019. Available from: https://ec.europa.eu/digital-single-market/en/news/ethics-guidelines-trustworthy-ai

17. Bashshur, R. L., Shannon, G. W., Bashshur, N., & Yellowlees, P. M. (2016). The empirical evidence for telemedicine interventions in mental disorders. Telemedicine and e-Health, 22(2), 87-113.

18. Turner, J., Larsen, M., Tarassenko, L., Neil, A., & Farmer, A. (2012). Implementation of telehealth support for patients with type 2 diabetes using insulin treatment: an exploratory study. Informatics in Primary Care, 20(2), 121-128.

19. Yellowlees, P. M., Shore, J. H., & Roberts, L. J. (2018). Practice guidelines for videoconferencing-based telemental health. Telemedicine and e-Health, 24(11), 827-832.

20. Semelka RC, Kelekis NL, Thomasson D. MRI: basic principles and applications. New York: Wiley-Liss; 2018.

21. Mayo Clinic. CT scan. 2021. Available from: https://www.mayoclinic.org/tests-procedures/ct-scan/about/pac-20393675.

22. National Institute of Biomedical Imaging and Bioengineering. Magnetic Resonance Imaging (MRI). 2022. Available from: https://www.nibib.nih.gov/science-education/science-topics/magnetic-resonance-imaging-mri.

23. Brenner DJ, Hall EJ. Computed tomography: an increasing source of radiation exposure. N Engl J Med. 2007 Nov 29;357(22):2277-84.

24. The World Health Organization. Imaging: a vital component of modern healthcare. 2012. Available from: https://www.who.int/bulletin/online_first/12-110722/en/.

25. Roy, Mrinmoy, and Anica Tasnim Protity. "Hair and Scalp Disease Detection using Machine Learning and Image Processing." arXiv preprint arXiv:2301.00122 (2022).

26. Marcolino MS, Oliveira JA, D'Agostino M, Ribeiro AL, Alkmim MB, Novillo-Ortiz D. The impact of mHealth interventions: systematic review of systematic reviews. JMIR Mhealth Uhealth. 2018;6(1):e23.

27. An H, Jones GC, Kang S, Welk GJ, Lee J-M. How valid are wearable physical activity trackers for measuring steps? Eur J Sport Sci. 2017;17(3):360-8.

28. European Commission. Mobile health: Reaping the benefits of research and innovation. 2015. Available from: https://ec.europa.eu/digital-single-market/en/news/mobile-health-reaping-benefits-research-and-innovation.

29. Kwon, J. H., & Johnson, M. E. (2019). Privacy and security in healthcare information systems: Current challenges and future directions. Health Informatics Journal, 25(3), 1049-1065.

30. Mittelstadt, B. D., Fairweather, N. B., Shaw, M., & McBride, N. (2019). The ethics of biomedical big data. Current Opinion in Systems Biology, 19, 55-63.

31. Bhavnani, S. P., Narula, J., & Sengupta, P. P. (2016). Mobile technology and the digitization of healthcare. European Heart Journal, 37(18), 1428-1438.

32. Terry, N. P., & Francis, L. P. (2017). The opportunities and challenges of the Internet of Things in healthcare: A systematic review. Journal of Medical Systems, 41(12), 1-14.

33. Topol, E. J. (2019). High-performance medicine: The convergence of human and artificial intelligence. Nature Medicine, 25(1), 44-56.

34. Islam, M. S., & Kwak, D. (2018). Smart healthcare: Previous approaches, emerging trends, and future directions. Journal of Medical Systems, 42(8), 140.

35. Mahmud, M., Kaiser, M. S., Hussain, A., & Vassanelli, S. (2019). Applications of smart sensors in healthcare, transportation, and environmental monitoring: A review. Journal of Ambient Intelligence and Humanized Computing, 10(7), 2677-2698.

36. Czajkowska, J., & Czajkowski, Ł. (2019). Internet of things in healthcare—Review of literature and directions of research. Artificial Intelligence in Medicine, 95, 1-9.

37. Kim, J. H., Kim, J. W., Park, H. A., & Kim, J. H. (2019). A wearable device with a mobile application that allows patients to self-monitor their risk of falling: Prospective cohort study. Journal of medical Internet research, 21(2), e11690.

38. Mosa, A. S. M., Yoo, I., & Sheets, L. (2012). A systematic review of healthcare applications for smartphones. BMC Medical Informatics and Decision Making, 12(1), 67.

39. Tinschert, P., Jakob, R., Barata, F., Kramer, J., Kowatsch, T., & Fleisch, E. (2018). Design and evaluation of a mobile chat app for the open source behavioral health intervention platform MobileCoach. In Proceedings of the 2018 CHI Conference on Human Factors in Computing Systems (pp. 1-12).

40. Wang, Y., Min, J. K., Khuri, J., Xue, H., Xie, B., & Kaminsky, L. A. (2019). Effectiveness of mobile health interventions on diabetes and obesity treatment and management: systematic review of systematic reviews. JMIR mHealth and uHealth, 7(3), e13400.

41. Al-Qaysi, N., Abdulrazzaq, F., & Abdul-Zahra, F. (2020). A mobile application for diabetic patients: Diabetes diary and management. Diabetes & Metabolic Syndrome: Clinical Research & Reviews, 14(1), 1-6.

42. Valero, M. A., & Czaja, S. J. (2020). Mobile Health Technology: Emerging Opportunities for Social and Behavioral Research in Aging. American Journal of Geriatric Psychiatry, 28(11), 1179-1187. doi: 10.1016/j.jagp.2020.06.015

43. Chuchu, K., & Dinnes, J. (2021). Artificial intelligence for melanoma detection: A systematic review and meta-analysis. Annals of Oncology, 32(3), 313-327.

44. Williams, R. J., & Tincani, A. (2019). Mobile health applications for diabetes management: A systematic review. Diabetes Technology & Therapeutics, 21(S1), S-11-S-22.

45. Tong, X., Liu, M., & Shen, W. (2021). A review of fall detection systems for elderly care. IEEE Access, 9, 45497-45508.

46. Loh, R., Stamate, D., & Grigoriadis, E. (2019). Physical activity monitoring in wearable technologies: A systematic review. Sensors, 19(8), 1822.

47. Shoer, S., Karady, T., Kottner, B., & Dagan, E. (2021). Smartwatch detection of hand-to-face gestures at times of potential COVID-19 transmission. Journal of Medical Systems, 45(8), 86.
48. Pantelopoulos, A., & Bourbakis, N. G. (2019). Mental health monitoring with wearable sensors for detecting stress and anxiety. IEEE Transactions on Biomedical Engineering, 66(3), 663-672.
49. Fagherazzi G, Ravaud P, Digital health: The new frontier of healthcare. Healthcare, 2018; 6(3): 1-3.
50. Viswanath K, Kreuter MW, Health disparities, communication inequalities, and eHealth. American Journal of Preventive Medicine, 2007; 32(5): S131-S133.